\documentclass[aps,10 pt,prd,twocolumn,nofootinbib,notitlepage,superscriptaddress]{revtex4-2}

\usepackage{graphicx}
\usepackage{dcolumn}
\usepackage{bm}
\usepackage{amsmath,amssymb}
\usepackage{multirow}
\usepackage{slashed}
\usepackage{xcolor}
\usepackage{physics}
\usepackage[colorlinks=true, pdfstartview=FitV, bookmarks=true, bookmarksnumbered=true, breaklinks]{hyperref}
\usepackage{mathtools,braket}
\usepackage{soul}
\usepackage{lipsum}  
\usepackage{color}
\definecolor{blue}{rgb}{0.0, 0.0, 1.0}
\definecolor{red}{rgb}{1.0, 0.0, 0.0}
\definecolor{royalblue}{rgb}{0.0, 0.14, 0.4}
\hypersetup{linkcolor=royalblue, citecolor=blue, urlcolor=royalblue}

\usepackage{hyperref}
\hypersetup{colorlinks=true,citecolor=blue,linkcolor=blue,urlcolor=blue}

\usepackage[mathlines]{lineno}
\usepackage{tikz,xcolor,hyperref}
\definecolor{lime}{HTML}{A6CE39}
\DeclareRobustCommand{\orcidicon}{%
	\begin{tikzpicture}
	\draw[lime, fill=lime] (0,0) 
	circle [radius=0.16] 
	node[white] {{\fontfamily{qag}\selectfont \tiny ID}};
	\draw[white, fill=white] (-0.0625,0.095) 
	circle [radius=0.007];
	\end{tikzpicture}
	\hspace{-2mm}
}

\foreach \x in {A, ..., Z}{%
	\expandafter\xdef\csname orcid\x\endcsname{\noexpand\href{https://orcid.org/\csname orcidauthor\x\endcsname}{\noexpand\orcidicon}}
}

\begin{document}

\title{Effects of flavor-mixings on charged kaon and pion parton distribution functions}

\author{Fabio L. Braghin\orcidC}
\email[E-Mail:]{braghin@ufg.br}
\affiliation{Instituto de Física, Federal University of Goiás, Avenida Esperança, s/n, 74690-900, Goiânia, GO, Brazil}

\author{Parada~T.~P.~Hutauruk\orcidA}
\email[E-Mail:]{phutauruk@hiroshima-u.ac.jp}
\affiliation{International Institute for Sustainability with Knotted Chiral Meta Matter (WPI-SKCM$^2$),Hiroshima University, 1-3-1 Kagamiyama, Higashi-Hiroshima, Hiroshima 739-8531, Japan}

\date{\today}

\begin{abstract}
We investigate the charged kaon and pion parton distribution functions (PDFs) in the U(3) Nambu–Jona-Lasinio model, considering a novel flavor-mixing interaction arising from vacuum polarization that differs from the instanton-induced 't Hooft interaction. In this work, the proper-time regularization scheme is employed, and, effectively, it might take quark confinement into account. The gap equations, the meson masses, and the meson–quark coupling constants are calculated in the presence of flavor-mixing interactions. The valence-quark distributions of the charged kaon and pion are calculated and compared with existing experimental data and JAM analysis results at scales $\mu^2 =$ 4 and 27 GeV$^2$. The strength of mixing effects on the PDFs is found to be proportional to the quark effective masses and their differences. We further find that the dominant flavor-mixing effects in the pion valence up-quark distribution at scales $\mu^2 =$ 27 and 4 GeV$^2$ are around $x \simeq 0.3$ and $x \simeq 0.8$, while for the kaon, the effects are shown at around $x \simeq 0.2$ and $x \simeq 0.7$. We also find similar strength of mixing effects on the PDFs at both scales $\mu^2 =4$ and 27 GeV$^2$; however, it certainly improves the pion and kaon PDFs. 
\end{abstract}

\maketitle

\section{ Introduction}
\label{sec:intro}
The parton distribution function (PDF) is an important quantity for investigating the internal structure of hadrons. A better understanding of the hadron PDF in terms of the quark and gluon degrees of freedom should improve our understanding of nonperturbative aspects of quantum chromodynamics (QCD)~\cite{Lorce:2025aqp,Gross:2022hyw}. Pions and kaons, as the lightest bound states of hadrons and pseudo-Goldstone bosons, provide a great avenue to study the QCD properties, in particular, nonperturbative QCD properties.
There are many calculations of the pion and kaon PDFs~\cite{Hutauruk:2016sug,Watanabe:2017pvl,Frederico:1994dx,Bentz:1999gx,Avila:2002xd,Nguyen:2011jy,Hutauruk:2025wkn,Hutauruk:2022zju,Chen:2024dhz,Aicher:2010cb} that have been made by considering the charge symmetry (CS), in which the current quark mass for different flavors is the same, e.g., $m_u = m_d$,  where $m_u$ and $m_d$ are the current quark masses for the up and down quarks, respectively,
(See reviews in Refs.~\cite{Miller:2006tv,Londergan:2009kj,Londergan:1998ai,Miller:1990iz}). In fact, CS is violated in nature, i.e., $m_u \neq m_d$, and this 
manifests in the QCD Lagrangian in terms of current quark masses, although electromagnetic effects also contribute~\cite{Miller:2006tv,GasserLeutwyler,Donoghue}.
Later, several studies have been performed to investigate such a charge symmetry breaking (CSB) or charge symmetry violation (CSV) effect in the effective theories~\cite{Wang:2015msk,Thomas:2014dxa,Hutauruk:2019jja,Sather:1991je} and lattice QCD simulations~\cite{Horsley:2010th,Shanahan:2015caa}.
Nevertheless, the CS limit might be useful to disentangle different effects taking place in the intricate quark dynamics inside hadrons.

The scarcity of experimental data for the pion and kaon PDFs, due to the lack of pion and kaon targets in the experiments, makes it difficult to fully understand their detailed structure at present. Charged pion and kaon PDFs are planned to be measured at the Electron-Ion Collider (EIC)~\cite{Arrington:2021biu}, the Electron-Ion Collider in China (EicC)~\cite{Anderle:2021wcy}, the JPARC~\cite{Sawada:2016mao}, and the AMBER/COMPASS++ at CERN~\cite{Adams:2018pwt} through the Sullivan~\cite{Lu:2025bnm} and Drell-Yan processes~\cite{Londergan:1994gr,Klest:2025fwx,Ding:2019lwe}, which prove to be an excellent way to probe the meson internal structure. These experiments are expected to provide new data on pion and kaon PDFs with higher accuracy, where our results of the present study can be potentially confirmed and verified by the data, helping to disentangle different aspects of
quark dynamics and meson structure.

Quark mixing, with consequences for quark mass differences, is usually known to emerge from instanton-induced flavor-mixing, which can be understood in quark dynamics by means of the 
't Hooft determinantal interaction~\cite{tHooft:1986ooh}. More recently, it has been shown that vacuum polarization or gluon exchanges can also induce flavor-mixing due to flavor symmetry breaking (FSB), which has been identified by means of 
flavor-dependent effective quark interactions~\cite{Braghin:2020yri,braghin2025POS}.
These mixing mechanisms, just mentioned above, have slightly different roles on quark masses and pseudoscalar meson masses (neutral meson mixings). The vacuum polarization (or gluon exchange) induced mechanism does not introduce any further energy scales or new free parameters. As a consequence, one is left with a set of self-consistent equations for constituent masses and coupling constants. Some effects of flavor-dependent effective quark interactions have been investigated in the light meson sector~\cite{Braghin:2020yri,Braghin_2022,braghin2025POS,Braghin:2022uih} and for some aspects of the heavier meson sector in two different models~\cite{pmwg-8t5d,EPJA-2023}.
These flavor-dependent interactions lead to subtle modifications in observables such as mixing interactions for sea and constituent quarks, a slightly improved description of observables, and also neutral meson mixing due to FSB. 
This approach leads to further consequences, such as, for example, the need for a single energy scale to describe flavor SU(4) pseudoscalar meson multiplet spectrum~\cite{pmwg-8t5d,EPJA-2023}.

In this work, we investigate some effects of a particular flavor-mixing, which arises from vacuum polarization, in the charged pion and kaon PDFs by considering the Nambu--Jona-Lasinio (NJL) model as well as the dynamical quark masses, the kaon and pion masses, and the meson-quark coupling constants. The NJL model is a successful low-energy QCD effective model that encompasses several features of the quark model for the meson sector~\cite{Vogl:1991qt,Klevansky:1992qe,Hatsuda:1994pi,Bijnens:1995ww}.
By means of dynamical chiral symmetry breaking, as a mechanism for the generation of hadron masses, the model produces many global meson observables in agreement with experimental results. 
More recently, effects from the FSB
have become a highly investigated topic in the search for a more precise description of meson spectra and properties.
As a low-energy effective model that is expected to be traced back to QCD, its parameters should all be expected to depend on quark mass differences, i.e., FSB and CSB.
Although we keep the CS limit, the strange-quark mass is large, as needed to reproduce kaon masses and decay constants, so that the resulting flavor-mixing 
appears. Besides observing the flavor-mixing effects on the dynamical quark masses, the meson masses, and the coupling constants, we investigate the effects of flavor-mixing on kaon and pion valence-quark distributions at scales $\mu^2 =$ 4 and 27 GeV$^2$, evolved using the next-to-leading order Dokshitzer–Gribov–Lipatov–Altarelli–Parisi (NLO-DGLAP) evolution equations~\cite{Miyama:1995bd}.

The present paper is organized as follows. In Sec.~\ref{sec:NJLmix}, the U(3) NJL model with global properties and the flavor-dependent interactions is briefly introduced and described. In Sec.~\ref{sec:pdf}, we then present the formulation of the leading twist-2 of the meson PDFs and the NLO-DGLAP evolution equations. Section~\ref{sec:result} presents our numerical results, first for the determination of the parameters of the model by reproducing global pion and kaon observables for different sets of model parameters with and without flavor-mixing. With these parameters, the meson PDFs are calculated and evolved to higher scales $\mu^2 =$ 4 and 27 GeV$^2$ by means of the NLO-DGLAP equations~\cite{Miyama:1995bd} to be comparable to experimental data~\cite{E615:1989bda} and the JAM global analysis~\cite{Barry:2021osv}.
In Sec.~\ref{sec:conlusion}, a summary of this work is presented.

\section{NJL model with flavor-mixings}
\label{sec:NJLmix}
In this section, we describe the U(3) NJL model with the flavor-mixing interactions
that have been extensively discussed in Refs.~\cite{Braghin:2020yri,Braghin_2022,Braghin:2022uih,braghin2025POS}. In the following, we outline how these interactions are captured in the NJL model before it is incorporated into the calculation of the pion and kaon PDFs. The generating functional of the model can be simply written as
\begin{eqnarray}
    \mathcal{Z} \big[ \bar{\eta}_f, \eta_f \big] &=& \mathcal{N} \int D \big[ \psi_f, \bar{\psi}_f \big] \exp i \int d^4 x \Big[  \mathcal{L}_{\mathrm{eff}} \big[\bar{\psi}_f \psi_f \big] \nonumber \\
    &+& \int d^4 x \big( \bar{\eta}_f \psi_f + \eta_f \bar{\psi}_f \big) \Big],
\end{eqnarray}
where $\mathcal{N}$ is an irrelevant normalization constant, and the quark flavor currents are defined as $\mathcal{J}_i^{\Gamma}=\bar{\psi}_q \lambda_i \Gamma \psi_q,
$ with $\Gamma$ denoting the Dirac structure and $\lambda_i$ denoting the flavor Gell-Mann matrices. The flavor indices $f,f_1,f_2,\ldots = u,d,s$ belong to the fundamental representation, and $i,j = 0,1,\ldots, N_f^{2-1}$ denote flavor indices in the adjoint representation.
The corresponding Lagrangian density is given by
\begin{eqnarray}
\label{eq1}
\mathcal{L}_{\mathrm{eff}} \big[ \bar{\psi}_f,\psi_f \big] &=&  \bar{\psi}_f \big( i \slashed{\partial} - m_f \big) \psi_f + G_{ij} \big[\mathcal{J}_{i}^{\Gamma} \mathcal{J}_j^{\Gamma} \big].
\end{eqnarray}
 The parameters $m_f$ correspond to the current (bare) quark masses, whereas $G_{ij}$ are the effective coupling constants. In the limit of degenerate quark masses, these couplings reduce to the standard coupling constant $G_0$ of the original NJL model. The effective couplings can be, initially, written as
 \begin{eqnarray}
G_{ij} = G_0 + \Delta G_{ij},
 \end{eqnarray}
where $G_0$ is the coupling constant of the standard NJL model, while $\Delta G_{ij}$ denotes the vacuum-polarization corrections. In general, the effective couplings in the scalar and pseudoscalar channels differ, and this difference is proportional to the current or constituent quark masses, reflecting explicit chiral-symmetry breaking. In the present work, following Ref.~\cite{Braghin:2020yri}, we adopt a common effective coupling for both the scalar and pseudoscalar channels and consider the pseudoscalar-channel coupling. Using the notation $\int_k \equiv \int \frac{d^4k}{(2\pi)^4}$, the effective coupling constants can be given by
\begin{eqnarray} 
\label{G2}
\Delta {G}_{ij} &=&  \frac{N_c G_0^2}{2} \int_k
\mathrm{Tr}_{D,F} \big[ S_f (k)   i \gamma_5\lambda_i  S_g (k)   i \gamma_5 \lambda_j \big],
\end{eqnarray}
where $S_f(k)$ are the quark propagators with flavor $f$ in terms of the quark effective masses extracted from the gap equations. Besides the diagonal couplings $G_{ii}$, which determine the meson flavor eigenstates, off-diagonal couplings $G_{ij}$ $(i\neq j), (i,j=0,3,8)$ are also present and generate 
explicit mixing among flavor-neutral currents, which are not considered in the present work.

The flavor-dependent corrections to the coupling constants, denoted by $\Delta G_{ij}$, can be expressed in terms of momentum-space integrals associated with the corresponding quark flavor content as 
\begin{eqnarray}
\Delta  G_{11} &=& I_{ud} + I_{du}, \\
\Delta  G_{33} &=&   I_{uu} + I_{dd}, \\
\Delta  G_{44} &=& I_{us} + I_{su}, \\
\Delta G_{66} &=& I_{ds} + I_{sd}, \\
\Delta G_{88} &=& \frac{1}{3} ( I_{uu} + I_{dd} + 4  I_{ss} ),\\
\Delta G_{00} &=& 
\frac{2}{3} ( I_{uu} + I_{dd} + I_{ss} ), \\
\Delta G_{08} &=&  \frac{ \sqrt{2} }{3} 
 ( I_{uu} + I_{dd} - 2  I_{ss} ), \\
\Delta G_{38} &=&  \frac{1}{ \sqrt{3} } 
( I_{uu} - I_{dd} ), \\
\Delta  G_{03} &=&  \sqrt{ \frac{2}{3} }
( I_{uu} - I_{dd} ),
\end{eqnarray}
where the divergence in the momentum integrals, solved in the proper-time regularization scheme with the ultraviolet cutoff $\Lambda_{\mathrm{UV}}$, is expressed by
\begin{eqnarray}
I_{fg} &=& I_{ff} + I_{gg} - B ( M_f - M_g )^2 \int_{1/\Lambda_{\rm{UV}}^2}^{1/\Lambda_{\rm{IR}}^2} 
\frac{d \tau}{\tau} \nonumber \\
&\times& \int_ 0^1  d x \;  \exp \big[- \tau \big( \Delta_1 \big) \big],  \\
I_{ff} &=& B \Bigg[\Lambda_{\rm{UV}}^2 e^{ - M_f^2/\Lambda_{\rm{UV}}} \nonumber \\
&+& M_f^2  \int_{1/\Lambda_{\rm{UV}}^2}^\infty d \tau \frac{ e^{-\tau M_f^2} }{\tau} \Bigg], 
\end{eqnarray}
where $\Delta_1 = (M_f^2+M_g^2)/2 
- x (M_f^2 - M_g^2)$, $B = \frac{ N_c G_0^2 }{ 4 \pi^2} $, and $M_f$ stands for the constituent (dynamical) quark masses determined from the gap equations. Because the effective couplings depend on the constituent quark masses, while the latter are themselves obtained self-consistently from the interaction, a normalization condition must be imposed. We fix the normalization by choosing the charged-pion coupling $G_{11}$ as the reference value, such that the normalized coupling constants can be written as
\begin{eqnarray}
\label{normalization}
G_{ii} =  G_{0} 
\left[ \frac{ G_0 +  \Delta G_{ii}}{ G_0 + \Delta G_{11}} \right].
\end{eqnarray}
This ensures that any modifications of the charged pion mass may be exclusively due to the modifications of the quark effective masses in the gap equations. Other normalization prescriptions can be explored in future studies. Substituting the effective coupling constants into the effective action allows the quark-flavor structure of the corresponding flavor currents to be expressed as
\begin{eqnarray}
\label{GijGfg}
G_{ij} \mathcal{J}_i \mathcal{J}_j &=& 2 G_{f_1 f_2} \mathcal{J}_{f_1} \mathcal{J}_{f_2}.
\end{eqnarray}
This equation can be rewritten by neglecting  all the explicit nondiagonal mixing interactions $G_{i\neq j}$ or $G_{f\neq g}$, and one has
\begin{eqnarray} 
\label{G-K}
 2 G_{uu} &=&
 2  \frac{ G_{00}  }{3}
 + G_{33}  + \frac{G_{88} }{3} ,
\\
\label{G-K2}
2 G_{dd} &=& 2 \frac{ G_{00} }{3} + G_{33}  + \frac{G_{88} }{3}, 
\\
\label{G-K3}
2 G_{ss} &=& 2 \frac{ G_{00} }{3}
+ 4   \frac{G_{88} }{3} . 
\end{eqnarray}
Note that by neglecting  $G_{i\neq j}$, the resulting diagonal quark couplings $G_{ff}$ in Eq.~\eqref{G-K} acquire a dependence on the constituent masses of the other quark flavors, $M_{f_1}$ with $f_1 \neq f$. This mechanism is the so-called \textit{implicit flavor-mixing effect}.

\subsection{ Dynamical quark masses}
The gap equations corresponding to the flavor-dependent interactions in the fundamental representation can be derived from Eq.~(\ref{GijGfg}). In the absence of explicit mixing interactions, it yields
\begin{eqnarray}
M_f  &=&  m_f\;  - i \;  
\; G_{ff}\; \mathrm{Tr}_{D,C,k} \; D_f(k) ,
\end{eqnarray}
where $D_f(k)$ is the dressed quark propagator for flavor $f$, and $\mathrm{Tr}$ denotes the trace over color and Dirac indices together with the momentum integration. In contrast, the off-diagonal flavor couplings $G_{f\neq g}$ 
are subleading and must be considered together with the flavor-mixing terms generated by the instanton-induced 't Hooft determinant interaction~\cite{tHooft:1986ooh,Vogl:1991qt,Klevansky:1992qe,Hatsuda:1994pi,Bijnens:1995ww,Crewther:1978kq}. Throughout this work, as discussed above, such mixing effects are neglected, corresponding to the approximation $G_{f\neq g}=0$.

\subsection{Bethe-Salpeter Equations} 
At the Born level, the Bethe-Salpeter equation (BSE) corresponding to the effective action in Eq.~\eqref{eq1} can be written, by neglecting explicit mixing terms unneeded for the charged mesons, as
\begin{eqnarray}
1 = - G_{ii} \Pi_{ii}^{(f_1f_2)}.
\end{eqnarray} 
The corresponding BSEs, together with the self-consistent gap equations involving $G_{ff}$, then have the form
\begin{eqnarray} 
\label{BSE-Gff-ii}
&& (M_{ps}^2 -  ({M_{f_1}^*}  - {M_{f_2}^*}) ^2 )
\; G_{ij} 
\;\;
I_2^{f_1f_2} (Q^2) \nonumber \\
&=&
\frac{G_{ij}}{2}  \left( 
\frac{m_{f_1}}{ \bar{G}_{f_1f_1} M_{f_1}^* }
+ \frac{m_{f_2}}{ \bar{G}_{f_2f_2} M_{f_2}^* }
\right) \nonumber \\
&-&  \frac{1}{2}
\left( \frac{G_{ij}}{\bar{G}_{f_1f_1} } + 
\frac{G_{ij}}{\bar{G}_{f_2f_2} } \right) +  1.
\end{eqnarray}
The loop integrals are evaluated within the proper-time regularization scheme by introducing ultraviolet and infrared regulators, $\Lambda_{\mathrm{UV}}$ and $\Lambda_{\mathrm{IR}}$, respectively. The regularized momentum integral is then given by
\begin{eqnarray}
 I_2^{f_1f_2} (M_{ps}^2)
&=&
\frac{3}{\pi^2} \int_0^1 \; d \; x \; \int_{1/\Lambda_{\rm{UV}}^2}^{1/\Lambda_{\rm{IR}}^2}
\frac{ d\tau}{\tau} \nonumber \\
&\times&
e^{- \tau \left[
x (1-x) M_{ps}^2 + x M_{f_1}^2 + (1-x) M_{f_2}^2 \right]}.
\end{eqnarray}
The quark coupling constants $\bar{G}_{ff}$ are those entering the gap equations, which may be renormalized in a different way from the BSE.
In the present work, we allow for the implicit flavor-mixing with $\bar{G}_{ff} = G_{ff}$. Below, we explain the determination of the coupling $ G _ {ij} $.

\subsection{Calculations with \texorpdfstring{$G_{ij}$}{Gij}}
The calculations of the flavor coupling constants $G_{ij}$ are evaluated as follows:
\begin{itemize}
\item Standard gap equations of the model  with $G_0$ are solved, and then, self-consistently, $G_{ij}$ are computed.
\item Normalization of the coupling constants in Eq.~\eqref{normalization} must be implemented in the self-consistent determination of masses and coupling constants.
\item Parameters of the model were 
fitted in the standard way of the model by reproducing pion and kaon masses, for the original model with $G_0$.
\item
The effect of the flavor coupling constants on meson masses and other observables was verified by solving BSE and other equations
with the calculated flavor-dependent coupling constants.
\end{itemize}

Using the calculation steps explained above for the isospin symmetric case, the results for the dynamical up- and strange-quark masses, the meson masses, and the diagonal coupling constants for different parameter sets are tabulated in Table~\ref{tab1}, whereas the results for the pion-quark and kaon-quark coupling constants, extracted from the BSE, are given in Table~\ref{tab2}. Note that we are mostly interested in the quantitative effect of the implicit mixing on observables. Therefore, we have not redefined the parameters of the model, with flavor-dependent interactions, to fit all the observables, such as the meson masses.

\section{Meson parton distribution function}
\label{sec:pdf}
We begin by briefly recalling the generic definition of the twist-2 quark distribution functions on the lightcone, which is given as
\begin{eqnarray}
    \label{eq:NJL-gluon9}
    q_{\mathrm{ps}} (x) &=& \frac{p^+}{2\pi} \int d\xi^- \exp \big[ ixp^+ \xi^-\big] \nonumber \\
    &\times& \big< ps \mid \bar{\psi}_q (0) \gamma^+ \psi_q (\xi^-) \mid ps \big>_{c},
\end{eqnarray}
where $ps=(\pi,K)$ represents a pseudoscalar meson. 
The Bjorken scaling variable ($x$), interpreted as the longitudinal momentum fraction carried by the struck quark, is defined as $x=k^+/p^+$, with $k^+$ and $p^+$ denoting the plus-components of the quark and meson momenta, respectively. The parameters $\xi$ and $c$ correspond to the skewness variable and the connected matrix element. Employing the framework developed in Ref.~\cite{Hutauruk:2016sug}, the pion and kaon valence-quark distribution functions are evaluated from the two dominant Feynman diagrams presented in Fig.~6 of Ref.~\cite{Hutauruk:2016sug}. The corresponding valence-quark distributions can be written as
\begin{eqnarray}
    \label{eq:NJL-gluon10}
    q_{\mathrm{ps}} (x) &=& i  \int \frac{d^4k}{(2\pi)^4} \delta \big( k^+ - xp^+\big) \nonumber \\
    &\times& \mathrm{Tr} \big[ \bar{\Gamma}\, S_{l} (k) \gamma^+ \hat{P}_{u/d} S_{l} (k) \Gamma\, S_{s} (k-p) \big],  \\
     \bar{q}_{\mathrm{ps}} (x) &=& -i \int \frac{d^4k}{(2\pi)^4} \delta \big( k^+ + xp^+\big) \nonumber \\
     &\times& \mathrm{Tr} \big[ \Gamma\, S_{l} (k) \gamma^+ \hat{P}_{\bar{d}/\bar{s}} S_{s} (k) \bar{\Gamma}\, S_{s} (k+p) \big], 
\end{eqnarray}
where $\Gamma = g_{\mathrm{ps} q\bar{q}} i \gamma_5 \sum_{f_1 f_1} C_{f_1f_1} \big| f_1\big> \big< f_1 \big|$; the trace is performed over color, flavor, and Dirac indices, $g_{\mathrm{ps}q\bar{q}}$ is the meson-quark coupling constant; and $l$ refers to the light (up or down) quarks. The variable $\hat{P}_{u/d} = \big[ (2/3) \mathbf{1} \pm \lambda_3 + (1/\sqrt{3}) \lambda_8 \big]/2$ is the projection operator for the $u$ and $d$ quarks, while for the strange quark, it can be given by $\hat{P}_s = \big[ (1/3) \mathbf{1} - (1/\sqrt{3}) \lambda_8 \big]$.
We further make use of the relation $\bar{q}(x)=-q(-x)$. 
The pion and kaon valence-quark distribution functions are determined from their Mellin moments,
\begin{eqnarray}
    \mathcal{A}^n &=& \int_0^1 dx \, \, x^{n-1} q_{(\pi,K)}(x),
\end{eqnarray}
with superscript of $n$ being an integer. By employing the Ward-Takahashi identity, $S_q(k)\gamma^+S_q(k)=-\frac{\partial S_q(k)}{\partial k_+}$,
and performing the loop integrations using Feynman parameterization, one obtains explicit expressions for the kaon valence-quark and valence-antiquark distribution functions in the proper-time regularization scheme (See details in Ref.~\cite{Hutauruk:2016sug}). The final expression of the quark distributions of the meson can be written as
\begin{eqnarray}
    \label{eq:NJL-gluon11}
q_{\mathrm{ps}} (x) &=& \frac{3 g_{\mathrm{ps} q\bar{q}}^2}{4\pi^2} \int_{\tau^2_{\mathrm{UV}}}^{\tau^2_{\mathrm{IR}}} d\tau \exp \big[ -\tau \big( \Delta_1 \big)\big] \nonumber \\
&\times& \big[ \frac{1}{\tau} + x(1-x)(m_{ps}^2 - (M_l-M_s)^2)\big], \\
\bar{q}_{\mathrm{ps}} (x) &=& \frac{3 g_{\mathrm{ps} q\bar{q}}^2}{4\pi^2} \int_{\tau^2_{\mathrm{UV}}}^{\tau^2_{\mathrm{IR}}} d\tau \exp \big[ -\tau \big(\Delta_2 \big)\big] \nonumber \\
&\times& \big[ \frac{1}{\tau} + x(1-x)(m_{ps}^2 - (M_l-M_s)^2)\big],
\end{eqnarray}
where the variables $\Delta_1$ and $\Delta_2$ are, respectively, given by
\begin{eqnarray}
    \Delta_1 &=& x(x-1)m_{ps}^2 + xM_s^2 + (1-x) M_l^2, \\
    \Delta_2 &=& x(x-1)m_{ps}^2 + xM_l^2 + (1-x) M_s^2.
\end{eqnarray}
By construction, the valence-quark distribution functions of the pseudoscalar mesons satisfy the baryon-number and momentum sum rules. For instance, for the charged kaon, one has
\begin{eqnarray}
    \label{sum-rules}
    \int_0^1 dx u_v^{ps} (x) = \int_0^1 dx \bar{s}_v^{ps}(x) &=& 1 ,  \\
    \int_0^1 dx x \big[ u_{ps}^+ (x) + \bar{s}^+_{ps} (x)\big] &=& 1, 
\end{eqnarray}
where
$
u_v^{ps}(x)=u_{ps}(x)-\bar{u}_{ps}(x), \qquad
\bar{s}_v^{ps}(x)=\bar{s}_{ps}(x)-s_{ps}(x),
$
and
$
u_{ps}^+(x)=u_{ps}(x)+\bar{u}_{ps}(x), \qquad
\bar{s}_{ps}^+(x)=\bar{s}_{ps}(x)+s_{ps}(x).
$
Analogously, these definitions apply to the pion distributions. We emphasize that, at the initial NJL-model scale $\mu_0^2$, gluon distributions are absent because the gluonic degrees of freedom have been integrated out and absorbed into the effective coupling constants. Consequently, the NJL model contains no dynamical gluons at the hadronic scale. In the present work, the pion and kaon gluon distributions are generated dynamically through perturbative QCD evolution. Starting from the valence-quark distributions at the model scale $\mu_0^2$, the gluon and sea-quark distributions are radiatively generated via the NLO-DGLAP evolution equations, which form a coupled set of integro-differential equations.

Within perturbative QCD, the nonsinglet quark distribution satisfies the evolution equation
\begin{eqnarray}
    \label{eq:NJL-gluon14}
    \frac{\partial q_{\mathrm{NS}} (x,\mu^2)}{\partial \ln \big( \mu^2\big)} = P_{qq} (x, \alpha_s \big( \mu^2 \big)) \otimes q_{\mathrm{NS}}(x,\mu^2),
\end{eqnarray}
where $P_{qq}$ is the splitting function kernels. The convolution product expression of the splitting functions and the nonsinglet quark distribution in Eq.~(\ref{eq:NJL-gluon14}) is defined by
\begin{eqnarray}
    \label{eq:NJL-gluon15}
    P_{qq} (x, \alpha_s (\mu^2)) \otimes q_{\mathrm{NS}} (x,\mu^2) = \int _{x}^1 \frac{dz}{x} P \Big(\frac{x}{z} \Big) q_{\mathrm{NS}} (z,\mu^2). \nonumber \\
\end{eqnarray}

For another type of quark distribution (singlet quark distribution), it gives
$  q_{\mathrm{S}} (x) = \sum_i q_i^+ = \sum_i 
 (q_i (x) + \bar{q}_i (x)) $
with $i$ being the quark flavor. In the QCD evolution, the singlet quark distributions can be defined by
\begin{eqnarray}
    \label{eq:NJL-gluon17}
    \frac{\partial}{\partial \ln (\mu^2) } \Bigg[ \begin{matrix}
        q_S (x, \mu^2)  \\ g(x,\mu^2) 
    \end{matrix} \Bigg] = \Bigg[ \begin{matrix}
        P_{qq} & P_{qg} 
\\ P_{gq} & P_{gg}
\end{matrix}\Bigg] \otimes \Bigg[ \begin{matrix}
    q_{s} (x, \mu^2) \\ g(x,\mu^2) 
\end{matrix}\Bigg]. \nonumber \\
\end{eqnarray}

The meson gluon distributions are obtained by numerically solving the evolution equation given in Eq.~(\ref{eq:NJL-gluon17}). In perturbative QCD, the splitting functions entering the evolution equations admit a perturbative expansion in powers of the running coupling $\alpha_s(\mu^2)$, such that
\begin{eqnarray}
    \label{eq:NJL-gluon18}
    P (z,\mu^2) = \Bigg[ \frac{\alpha_s}{2\pi}\Bigg] P^{(0)} (z) + \Bigg[ \frac{\alpha_s}{2\pi}\Bigg]^2 P^{(1)} (z) + \cdot \cdot \cdot.
\end{eqnarray}
The perturbative expansion in Eq.~(\ref{eq:NJL-gluon18}) consists of the leading-order (LO) and next-to-leading-order (NLO) contributions, represented by the first and second terms, respectively. The corresponding NLO running coupling $\alpha_s(\mu^2)$ is given by
\begin{eqnarray}
    \label{eq:NJL-gluon19}
    \alpha_{s} = \beta_3 \Bigg[ 1 - \frac{\beta_1}{\beta_0^2} \frac{\ln \ln (Q_\Lambda)}{\ln (Q_\Lambda) }\Bigg] + \mathcal{O} \Bigg( \frac{1}{\ln^2 (Q_\Lambda)}\Bigg).
\end{eqnarray}
In Eq.~(\ref{eq:NJL-gluon19}), the quantities are defined by $Q_\Lambda=\mu^2/\Lambda_{\mathrm{QCD}}^2$, $\beta_0=\frac{11}{3}N_c-\frac{2}{3}N_f$, $\beta_1=\frac{34}{3}N_c^2-\frac{10}{3}N_cN_f-2C_FN_f$, and $\beta_3 = (4\pi/\beta_0) (1/\ln(Q_\Lambda))$. The parameters $N_c$ and $N_f$ represent the numbers of colors and active quark flavors, respectively, and $C_F=4/3$ is the Casimir invariant in the fundamental representation. The QCD scale parameter $\Lambda_{\mathrm{QCD}}$ depends on the number of active flavors as well as on the renormalization scheme. The valence-quark distributions of the mesons are subsequently evolved according to the NLO-DGLAP equations introduced above.

\section{Results and Discussion}
\label{sec:result}
Here, we present our numerical results for the pion and kaon parton distribution functions with and without flavor-mixing interactions at scales $\mu^2=27~\mathrm{GeV}^2$ and $\mu^2=4~\mathrm{GeV}^2$. Both scales are chosen to correspond to the scales employed in the experiment~\cite{E615:1989bda} and lattice-QCD calculations~\cite{Fan:2021bcr,Salas-Chavira:2021wui,Holligan:2024umc,Lin:2020ssv}, respectively. The calculation of the meson PDFs requires the constituent (dynamical) quark masses, the meson masses, the flavor-dependent coupling constants, and the meson-quark coupling constants as input. In the standard NJL model, i.e., without flavor-mixing interactions (G19-Set1), the model parameters were adapted from previous works~\cite{Hutauruk:2018zfk,Hutauruk:2016sug,Braghin_2022} 
and they are completely written in Table~\ref{tab1}. Note that the set of parameters taken from Ref.~\cite{Braghin_2022} was adapted to the proper time regularization scheme so that the results of the meson masses are quite different.

Having fixed the parameters of the standard NJL model, we next extend the analysis by incorporating flavor-mixing interaction effects. The flavor-dependent interactions modify the gap equations and the Bethe-Salpeter equations that are used to determine the pion and kaon masses. In this work, as mentioned above, we determine the flavor-mixing parameters using the standard NJL parameter sets in Refs.~\cite{Hutauruk:2016sug,Hutauruk:2018zfk}
and~\cite{Braghin_2022}. Results needed for the computation of the pion and kaon PDFs are also written in Tables~\ref{tab1} and~\ref{tab2}. For a given set of values for the current quark masses, and IR and UV cutoffs with different values of $G_0$ as provided in Table~\ref{tab1}, we evaluate the quark and meson masses through the gap and BSE equations, respectively, with and without the implicit flavor-mixing effects taken into account, as described below:
\begin{itemize}
\item Set 1: Standard BSE-NJL model without mixing ~\cite{Hutauruk:2016sug,Braghin_2022},
\item Set 4: Implicit flavor mixing in both gap equation and BSE,  with $G_{ii}$  and $G_{ff}$, in the absence of explicit mixings (off-diagonal mixings).
\end{itemize}
\begin{table*}[t]
\begin{ruledtabular}
		\renewcommand{\arraystretch}{1.6}
\caption{Results for the dynamical quark masses, the pion and kaon masses, and the diagonal coupling constants of the flavor mixings for the chosen normalization parameters Sets 1 and 4, where $G_{11} = G_0$. All parameters are in units of MeV, except the $G_{44}$, $G_{uu}$, $G_{11}$, $G_0$, and $G_{ss}$ are in units of GeV$^{-2}$.} 
\begin{tabular}{ c |c  c c c c c c c | c  c |  c   c c }
 $ \rm{Set}/G_0$  &  Notation & $\Lambda_{\rm{UV}}/\Lambda_{\rm{IR}}$  & $M_u$ &  $M_d$ & $M_s$ & $m_{0u}$ &  $m_{0d}$ & $m_{0s}$  &  $m_\pi$ & $m_K$ & $G_{44}$ & $G_{uu}$ & $G_{ss}$  \\ \hline 
$\rm{Set}1/19.04$ & G19-Set1 & 645/240  & 400   & 400 & 611  & 16.4  & 16.4  & 356 &140 & 495 & 19.04 & 19.04 & 19.04 \\
$\rm{Set4}/19.04$ &G19-Set4 & 645/240 & 389  & 389 &  596  & 16.4  & 16.4  & 356 & 104 & 510 & 17.33 &  18.29  &  17.16 \\ \hline
$\rm{Set1}/10$ & G10-Set1 & 645/240 & 158  &  158 &  450 & 7.5 & 7.5 & 260 & 140 & 505 & 10 & 10  &  10 \\
$\rm{Set4}/10$  &G10-Set4 & 645/240 &   143  &  143 &  439 & 7.5 & 7.5 & 260 
&   121 & 510 & 9.30  & 9.70  & 9.25
\end{tabular}
\label{tab1}  
	\end{ruledtabular}
\end{table*}
\begin{table}[t]
\begin{ruledtabular}
		\renewcommand{\arraystretch}{1.4}
\caption{Results of the meson-quark coupling constants for chosen normalization with different parameter sets: Sets 1 and 4.} 
\begin{tabular}{c |c c c} 
 $\rm{Set}/G_0$ & Notation & $g_{K qq}$ & $g_{\pi qq}$ 
  \\ \hline 
$\rm{Set1}/19.04$  & G19-Set1 & 4.57 & 4.23  \\
$\rm{Set4}/19.04$ &G19-Set4 & 4.36 &  4.16  \\ \hline
$\rm{Set1}/10$ & G10-Set1 & 2.92 &  2.76 \\
$\rm{Set4}/10$ & G10-Set4  & 2.84  & 2.73  
\end{tabular}
\label{tab2}  
	\end{ruledtabular}
\end{table}
\begin{figure*}[ht]
\centering
\includegraphics[width=1.8\columnwidth]{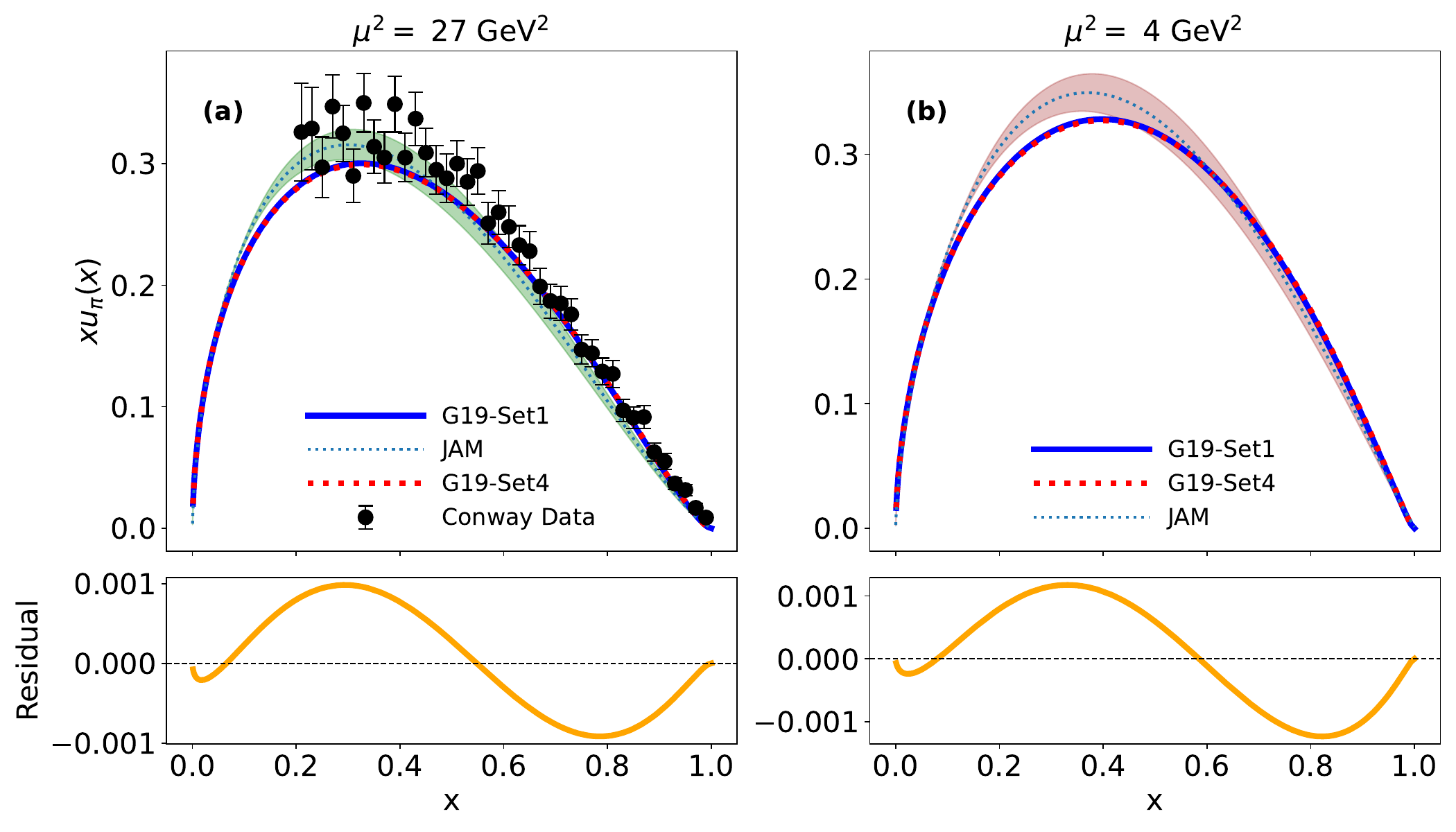} 
\caption{\label{fig1a} Valence quark distributions of the pion and their PDF differences at scales (a) $\mu^2 =$ 27  and (b) 4 GeV$^2$ as a function of the longitudinal momentum fraction $x$. Experimental data E-615 are taken from Ref.~\cite{E615:1989bda}. The pion valence quark distributions are evolved using the initial model scale $\mu_0^2 =$ 0.18 GeV$^2$. Note that the pion distribution is calculated using the $G_0 = 19.04$ GeV$^{-2}$ with G19-Set1 (without mixing) and G19-Set4 (with mixing), as shown in Table~\ref{tab1}.} 
\end{figure*}

Now we present our results for the pion and kaon PDFs at scales $\mu^2 =$ 4 and 27 GeV$^2$ for fixed $G_0 =$ 19.04 GeV$^{-2}$ with Set 1 and Set 4, as shown in Table~\ref{tab1}. These sets are represented by G19-Set1 and G19-Set4. Figure~\ref{fig1a}(a) shows the results of the valence up-quark distributions of the pion at scales $\mu^2 =$ 27  and 4 GeV$^2$. Also, their difference from the PDFs calculated in the standard NJL model is plotted below, named as "Residual". It shows that the implicit mixing leads to an asymmetric shift of the pion valence-quark distribution (orange solid line), which keeps the overall sum rules in Eq.~\eqref{sum-rules}. Also, it shows that the NJL-model valence-quark distribution of the pion with G19-Set1 (without mixing) at scale $\mu^2 =$ 27 GeV$^2$ is in good agreement with the existing experimental data~\cite{E615:1989bda}, which was measured at $\mu^2 =$ 27 GeV$^2$, and the JAM analysis~\cite{Barry:2021osv}, which confirms what was obtained in Ref.~\cite{Hutauruk:2016sug}. Note that we find that the effect of the implicit mixing in the pion PDFs is smaller than that of the effect of CSV~\cite{Hutauruk:2018zfk}, but they are of the same magnitude. In addition, we find that the flavor mixing effects certainly improve the pion PDF. High-precision data are needed for an assessment of such an effect.

In Fig.\ref{fig1a}(b), we show our results for the valence quark distribution of the pion at scale $\mu^2 =$ 4 GeV$^2$ for the same cases as Fig.~\ref{fig1a}(a). We find that the difference in the valence quark distribution for the pion between G19-Set1 and G19-Set4 has a similar behavior to that at scale
$\mu^2 = 27$GeV$^2$ and their differences are almost the same as those at scale $\mu^2 =$ 27 GeV$^2$. However, we find that the valence quark distribution of the pion at scale $\mu^2 =$ 27 GeV$^2$ is a bit smaller than that of scale $\mu^2 =$ 4 GeV$^2$, and this makes the relative contribution of the mixing slightly larger. In comparison to the JAM analysis results at scale $\mu^2 =$ 4 GeV$^2$, again, our results show good agreement with the JAM results.
\begin{figure*}[ht]
\centering
\includegraphics[width=1.8\columnwidth]{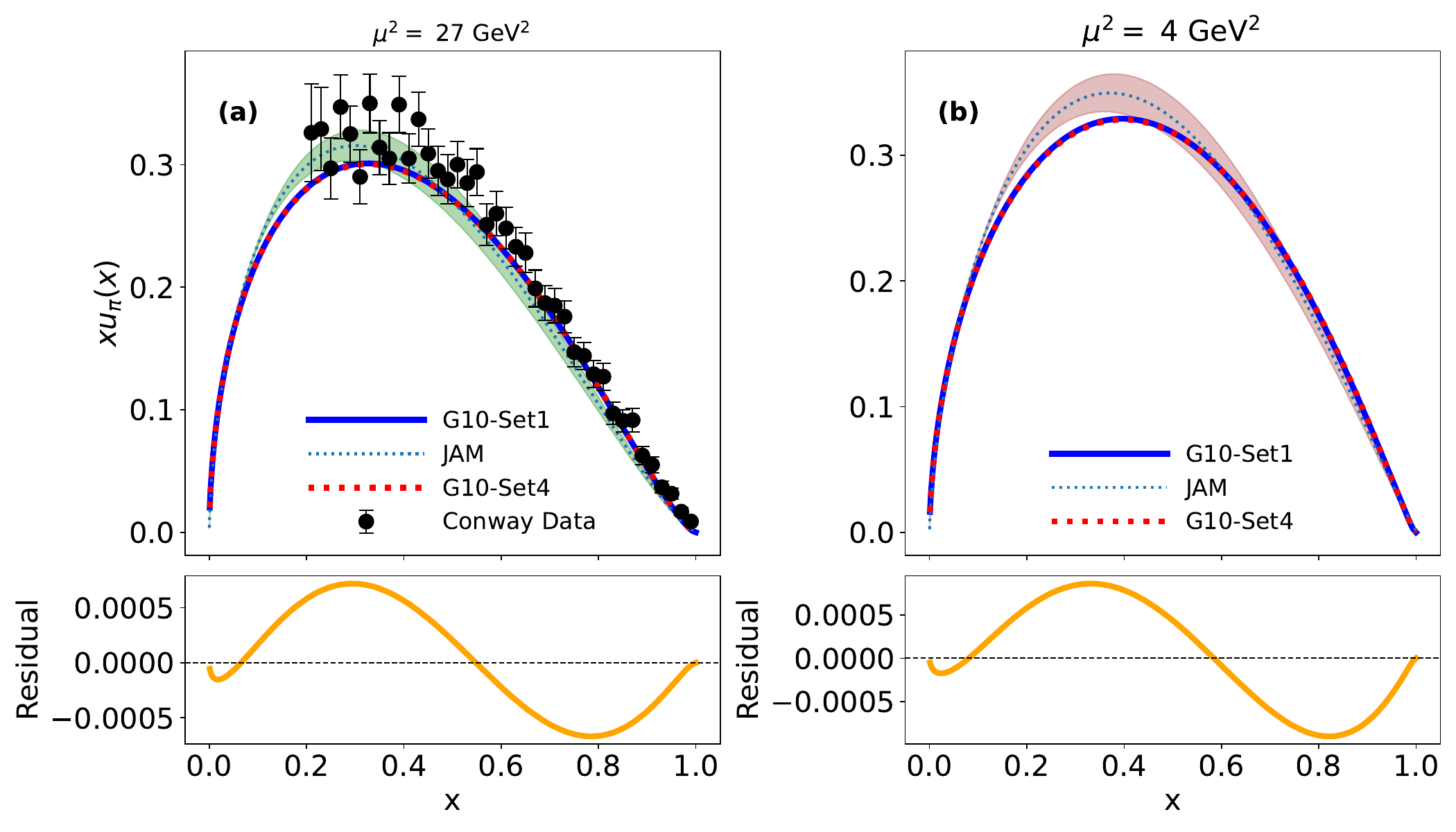}  
\caption{\label{fig1b} Same as in Fig.~\ref{fig1a}, but for $G_0 =$ 10 GeV$^{-2}$ with G10-Set1 and with G10-Set4.} 
\end{figure*}

Analogously, we also compute the valence quark for the pion at scales $\mu =$ 27 and 4 GeV$^2$ with fixed $G_0 =$ 10 GeV$^2$, which is adapted from Ref.~\cite{Braghin_2022} with the $\Lambda_{\mathrm{UV}} =$ 645 MeV and $\Lambda_{\mathrm{IR}} =$ 240 MeV, where these UV and IR cutoffs were used to calculate the valence quark in Fig.~\ref{fig1a}. Results for the valence quark distribution of the pion at scale $\mu^2 =$ 27 GeV$^2$ are depicted in Fig.~\ref{fig1b}(a). Again, the significant difference in the valence quark distribution of the pion at scale $\mu^2 =$ 27 GeV$^2$
is small. However, in comparison to differences of the valence quark distribution of the pion with fixed $G_0=$ 19.04 GeV$^{-2}$, the differences (modulus) for the $G_0 =$ 10 GeV$^{-2}$ are nearly $50 \%$  smaller at $x \simeq 0.3$ and at $x \simeq 0.8$. This implies that the flavor mixing significantly contributes to the pion PDFs in
specific $x$ regions.

The valence quark distribution of the pion at scale $\mu^2 =$ 4 GeV$^2$ with fixed $G_0 =$ 10 GeV$^{-2}$ is provided in Fig.~\ref{fig1b}(b). We find almost similar residuals of the valence quark distribution of the pion as in Fig.~\ref{fig1b}(a). In addition, the difference is also found in the magnitude of the valence quark distribution at scale $\mu^2 =$ 4 GeV$^2$ in comparison to that at scale $\mu^2 =$ 27 GeV$^2$.
\begin{figure*}[ht]
\centering
\includegraphics[width=1.8\columnwidth]{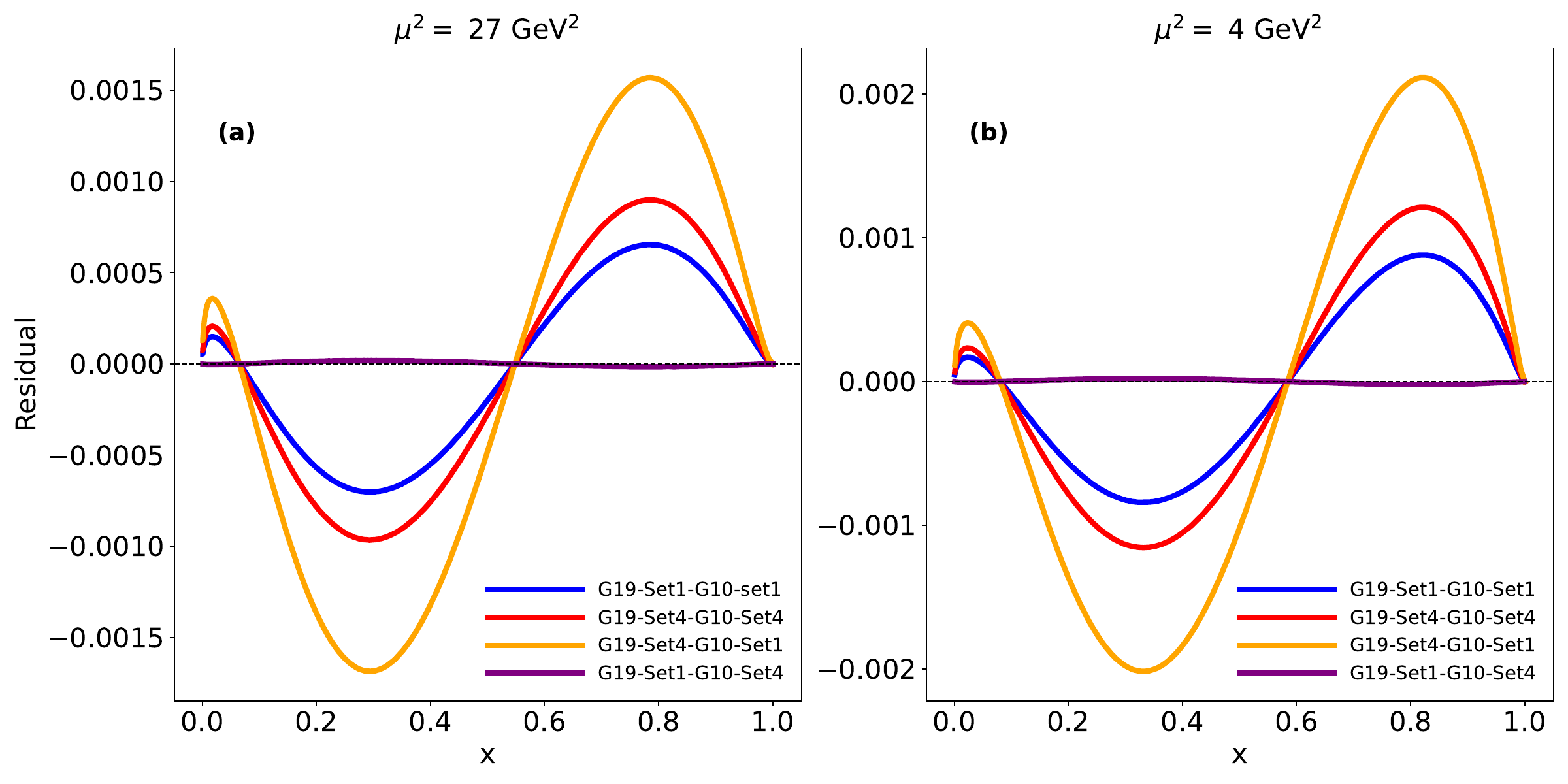} 
\caption{\label{fig1d} Differences of $xu_\pi (x)$ between fixed $G_0 =$ 19.04 GeV${-2}$ and $G_0 =$ 10 GeV$^{-2}$ at scales (a) $\mu^2 =$ 27 GeV$^2$ and (b) $\mu^2 =$ 4 GeV$^2$.} 
\end{figure*}

To see clearly the flavor mixing effect in the pion valence quark distribution of the pion at scales $\mu^2 =$ 27 GeV$^2$ ($\mu^2 =$ 4 GeV$^2$) for all corresponding parameter sets. The results of the difference of the pion valence quark distributions for (G19-Set1$-$G10-Set1), (G19-Set4$-$G10-Set4), (G19-Set4$-$G10-Set1), and (G19-Set1$-$G10-Set4) are given in Figs.~\ref{fig1d}(a) and \ref{fig1d}(b), respectively. 
It shows that G19-Set4$-$G10-Set1 gives the dominant flavor-mixing effects at $x \simeq 0.3$ and $x \simeq 0.8$, which are about 0.0015 and -0.0016, respectively, and others show similar behavior, but they are different in magnitude. 
However, there is a slight increase in the case when $\mu^2 =$ 4 GeV$^2$, as indicated in Fig.~\ref{fig1d}(b). 
\begin{figure*}[t]
\centering
\includegraphics[width=1.8\columnwidth]{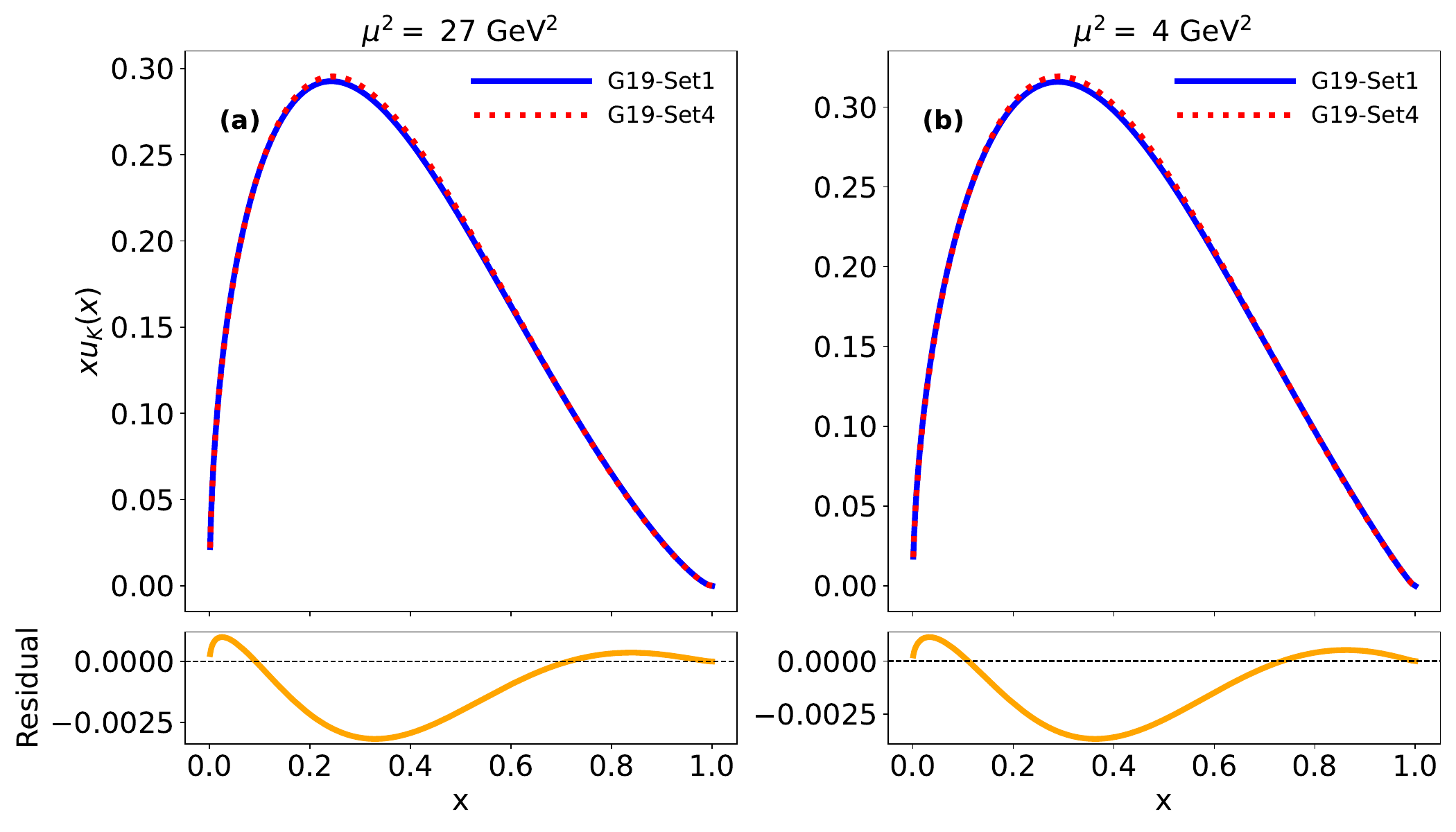} 
\caption{\label{fig2} Same as Fig.~\ref{fig1a} but for the up-quark distribution of the kaon.} 
\label{fig:kaon-up-19}
\end{figure*}

\begin{figure*}[t]
\centering
\includegraphics[width=1.8\columnwidth]{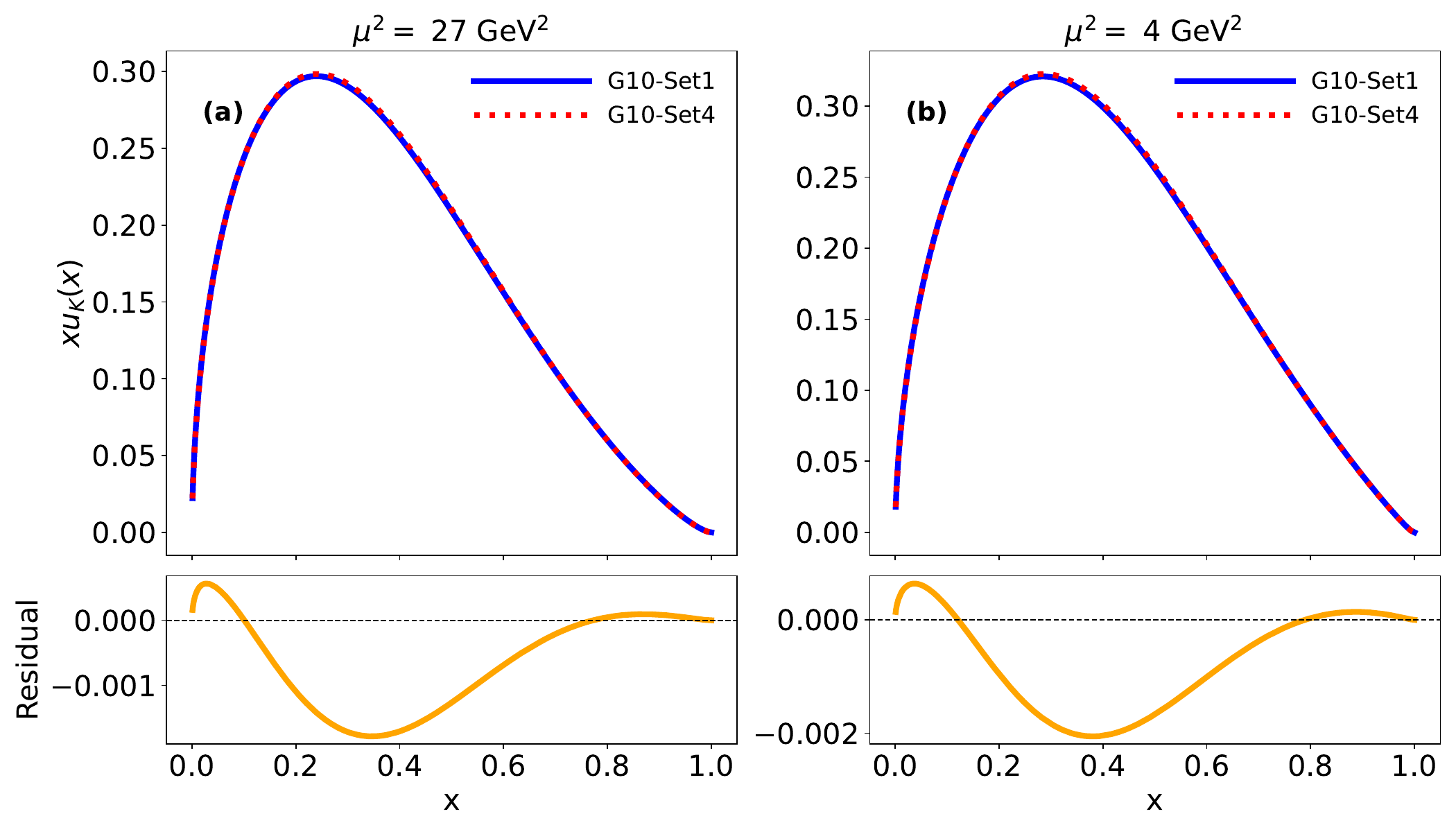} 
\caption{\label{fig2a} Same as Fig.~\ref{fig1b} but for the up-quark distribution of the kaon.} 
\label{fig:kaon-up-10}
\end{figure*}

Now, we investigate the kaon valence quark distribution at scales $\mu^2 =$ 27 and 4 GeV$^2$ for fixed $G_0 =$ 19.04 GeV$^{-2}$ (for fixed $G_0 =$ 10 GeV$^{-2}$) for Sets 1 and 4 in Fig.~\ref{fig:kaon-up-19}
(Fig.~\ref{fig:kaon-up-10}) (see details in Table~\ref{tab1}). It is found that the up quark in the kaon PDF plays a different role from the one in the pion, whose leading effect may be traced back to the large up and strange-quark mass differences that lead to the larger asymmetry in $x$.
Figure.~\ref{fig2}(a) shows that the results of the valence quark distribution of the kaon at scale $\mu^2 =$ 27 GeV$^2$ with fixed $G_0 =$ 19.04 GeV$^{-2}$. The kaon valence quark distribution is rather different from that for the pion. The dominant residual of the kaon valence quark distribution is given by 
Set 4 (orange solid line) at $x \simeq 0.3$, while the residual of the kaon valence quark distribution at $x \simeq 0.8$ is significantly suppressed in comparison to that for the pion. 

At scale $\mu^2 =$ 4 GeV$^2$, the residual result for the kaon valence quark distribution is even larger in comparison to that for the results obtained in Fig.~\ref{fig2}(a). Again, the kaon valence quark distribution at $x \simeq 0.8$ is suppressed, but at $x \simeq 0.3$, it is larger in comparison to that obtained in Fig.~\ref{fig2}(a).

The results for the kaon valence quark distribution and their residuals at scales $\mu^2 =$ 27 and 4 GeV$^2$ with fixed $G_0 =$ 10 GeV$^{-2}$ are depicted in Figs.~\ref{fig2a}(a) and (b), respectively. At scale $\mu^2 =$ 27 GeV$^2$, we find that the residuals of the kaon valence quark for the parameter set of G10Set1$-$G10Set4 are larger than those for others at around $x \simeq 0.3$, and are significantly suppressed at $x \simeq 0.8$. Again, a similar trend of the residuals of the kaon valence quark distribution at scale $\mu^2 = $ 4 GeV$^2$ is found, as shown in Fig.~\ref{fig2a}(b). Also, it shows that the magnitude of the residual increases as the scale increases to $\mu^2 =$ 4 GeV$^2$.
\begin{figure*}[t]
\centering
\includegraphics[width=1.8\columnwidth]{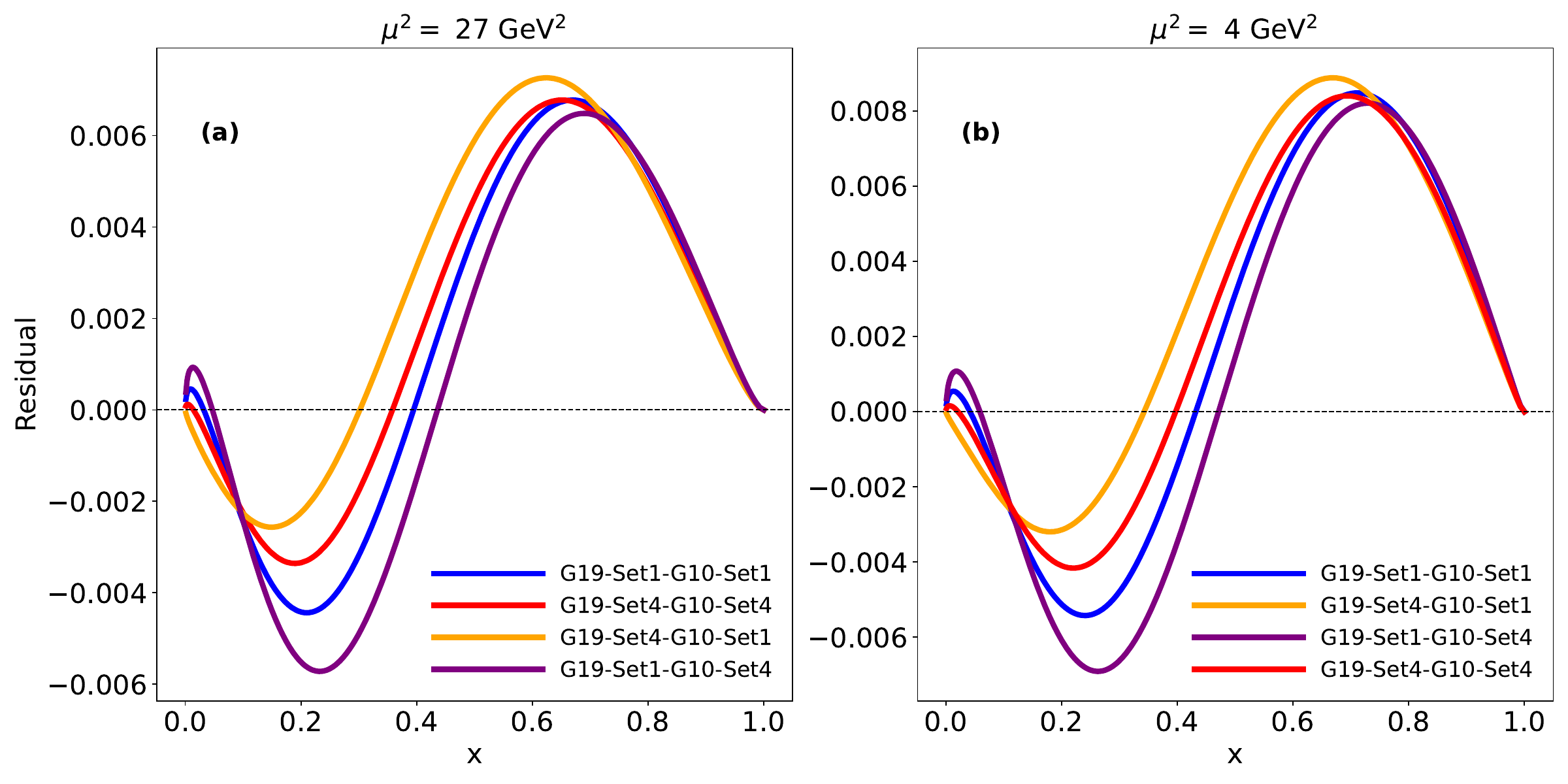} 
\caption{\label{fig2c} Same as Fig.~\ref{fig1d}, but for the up-quark distribution of the kaon.} 
\end{figure*}

Results for the differences of the valence quark distribution of the kaon at scales $\mu^2 =$ 27 GeV$^2$
and $\mu^2 =$ 4 GeV$^2$ for different parameter sets are given in Fig.~\ref{fig2c}(a) and (b). In Fig.~\ref{fig2c}(a), it shows that the dominant difference in the kaon valence quark distribution is given by G19-Set1$-$G10-Set4 (magenta solid line). However, it again clearly shows the flavor-mixing contributions to the kaon valence quark distribution. 
In Fig.~\ref{fig2c}(b), we show the difference of the kaon up valence quark distribution at scale $\mu^2 =$ 27 GeV$^2$ for fixed $G_0 =$ 10 GeV$^{-2}$. In comparison to the PDF difference result of Fig.~\ref{fig2c}(a), we find that the difference magnitude is a bit larger at around $x \simeq 0.2$ but is lower at around $x \simeq 0.7$. 

Results for the residuals of the kaon valence quark distribution at scale $\mu^2 =$ 4 GeV$^2$ are provided in Fig.~\ref{fig2c}(b). We find that the residuals between Fig.~\ref{fig2c}(a) and Fig.~\ref{fig2c}(b) have a similar shape and trend, but they have a slightly different magnitude at the corresponding values of $x$. Note that the PDF difference for the kaon up valence quark distribution is 1 order of magnitude larger than those for the pion.
\begin{figure*}[t]
\centering
\includegraphics[width=1.80\columnwidth]{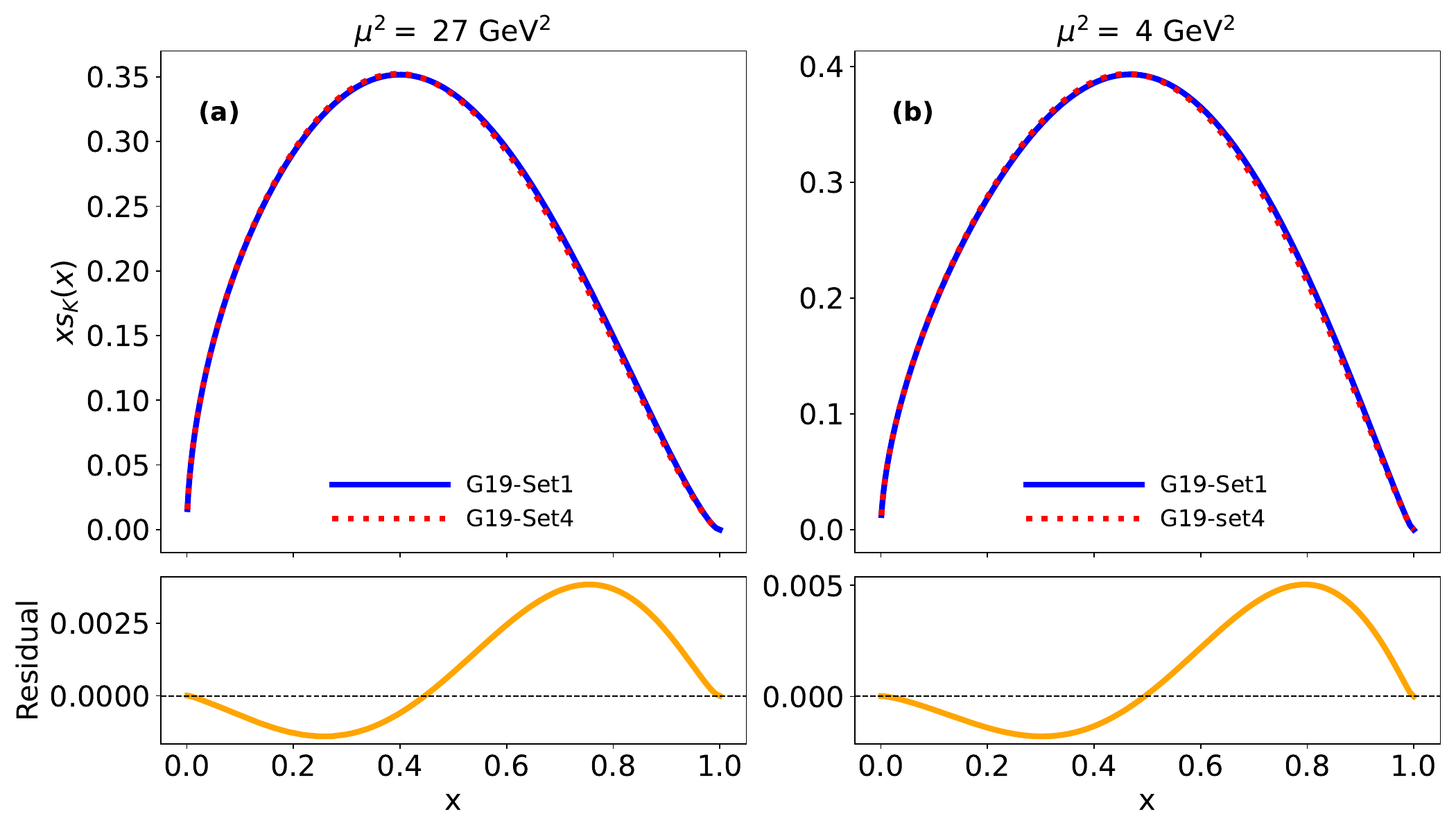}  
\caption{\label{fig3} Same as Fig.~\ref{fig1a} but for the antistrange-quark distribution of the kaon.} 
\end{figure*}

In addition to the kaon valence up-quark distribution, we also compute the valence antistrange-quark distribution of the kaon at scales $\mu^2 =$ 27 and 4 GeV$^2$ for fixed $G_0 =$ 19.04 GeV$^{-2}$ with Set 1 and Set 4, as given in Figs.~\ref{fig3}(a) and (b). We find different residual shapes and trends in comparison to those for the up-quark distribution of the kaon, as in Fig.~\ref{fig2}, where it shows to be suppressed at around $x \simeq 0.8$, while for the antistrange quark, it is suppressed at around $x \simeq 0.3$. 

Similar shapes and trends are shown for the kaon antistrange-quark distribution at scale $\mu^2 =$ 4 GeV$^2$ as shown in Fig.~\ref{fig3}(b), but it shows a bit larger magnitude in comparison to that obtained in Fig.~\ref{fig3}(a), as expected.
\begin{figure*}[t]
\centering
\includegraphics[width=1.8\columnwidth]{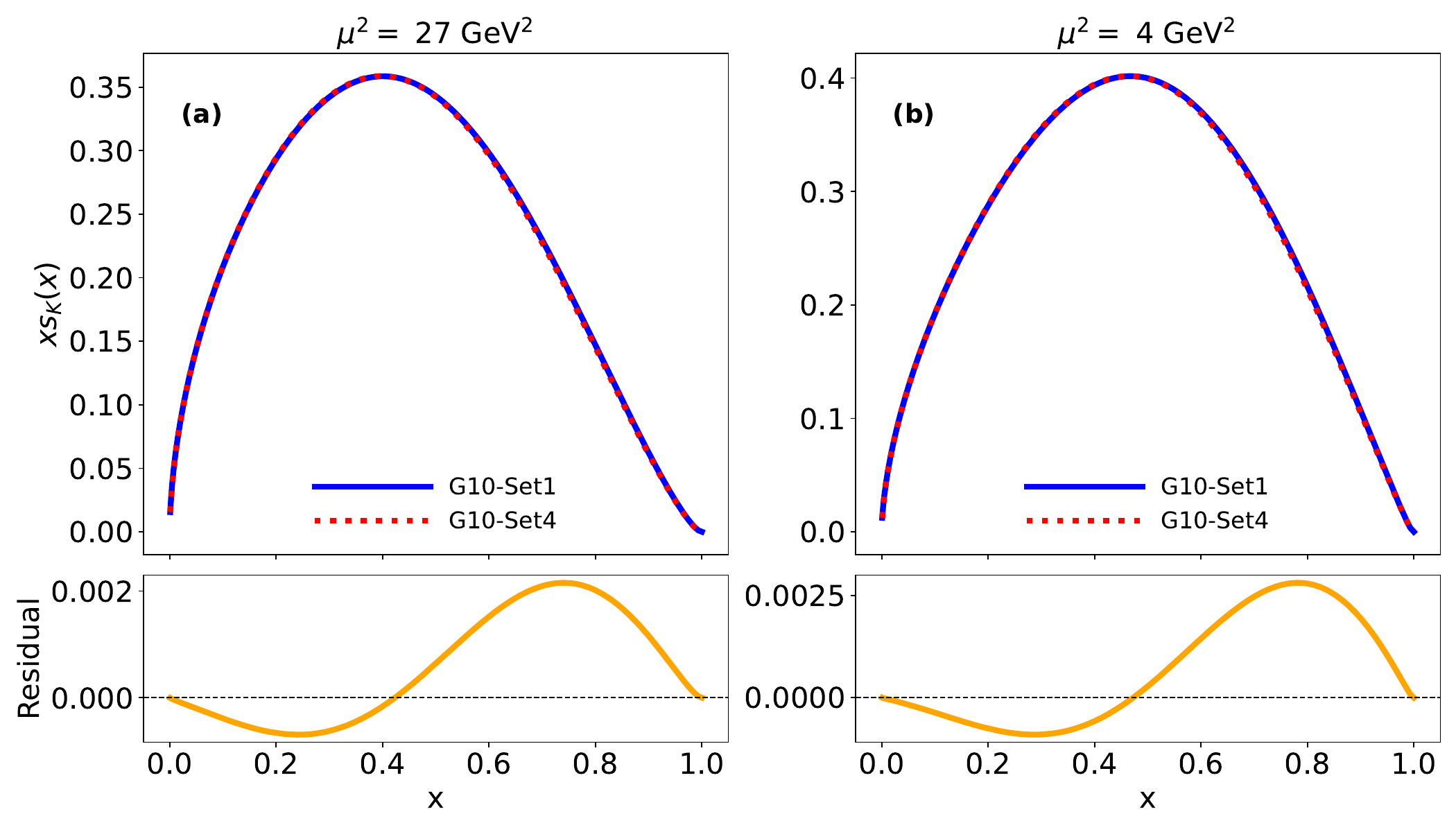}  
\caption{\label{fig3a} Same as Fig.~\ref{fig1b} but for the antistrange quark distribution of the kaon.} 
\end{figure*}

In Fig.~\ref{fig3a}, we show the results of the kaon valence antistrange-quark distribution and its residuals at scales $\mu^2 =$ 27 and 4 GeV$^2$ for fixed $G_0 =$ 10 GeV$^{-2}$. We find that the magnitude of the residual of the kaon valence antistrange-quark distribution is smaller than that obtained in Fig.~\ref{fig3}(a) at around $x \simeq 0.8$. However, the kaon valence antistrange-quark distribution for scale $\mu^2 =$ 4 GeV$^2$ at $x \simeq 0.8$ is larger than that of scale $\mu^2 =$ 27 GeV$^2$, as indicated in Fig.~\ref{fig3a}(b).
\begin{figure*}[t]
\centering
\includegraphics[width=1.80\columnwidth]{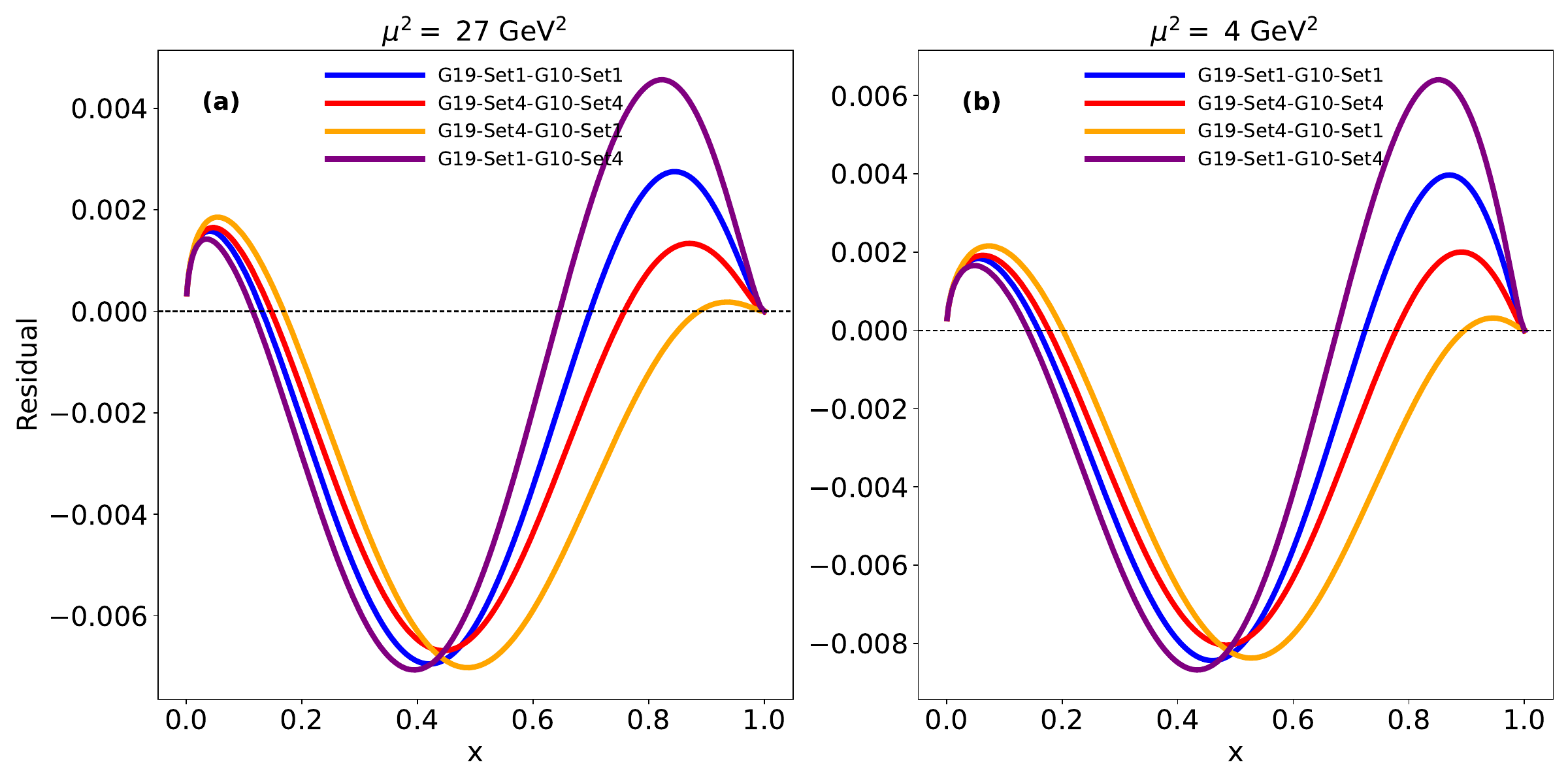}  
\caption{\label{fig3c} Same as Fig.~\ref{fig1d} but for the antistrange quark distribution of the kaon.} 
\end{figure*}

The difference results for the kaon valence antistrange-quark distribution at scales $\mu^2 =$ 27 and 4 GeV$^2$ are provided in Fig.~\ref{fig3c}. In Fig.~\ref{fig3c}(a), it is shown that the dominant difference is given by G19-Set1$-$G10-Set4 (magenta solid line) at around $x \simeq 0.4$ and $x \simeq 0.8$ in comparison to others.

Figure~\ref{fig3c}(b) shows that G19-Set1$-$G10-Set4 gives the larger difference values. The shapes and trend of the difference results are almost the same as those obtained in Fig.~\ref{fig3c}(a). At scale $\mu^2 =$ 4 GeV$^2$, the difference results of the kaon valence antiquark distribution in Fig.~\ref{fig3c} (b) have larger values in comparison to those obtained in Fig.~\ref{fig3c} (a), as expected.

\section{Conclusion}
\label{sec:conlusion}
To summarize, in the present work, we have investigated the flavor-mixing effects on the kaon and pion parton distribution functions in the U(3) NJL model, incorporating the proper-time regularization scheme for ultraviolet divergences and simulating quark confinement. To clearly investigate the flavor-mixing in the kaon and pion PDFs, we considered different parameter sets with and without flavor-mixing. With these parameters, we evaluated the pion and kaon PDFs at scales $\mu^2 =$ 27 and 4 GeV$^2$ using the NLO-DGLAP evolution equations~\cite{Miyama:1995bd}. 

In the analysis, we found that the pion valence quark distribution using the standard NJL model is consistent with the existing pion data~\cite{E615:1989bda} and JAM global QCD analysis~\cite{Barry:2021osv} at $\mu^2 =$ 27 GeV$^2$ at moderate and higher values of $x$. However, experimental data at lower $x$ are still absent, which is expected to be available in the future. On the other hand, we found that the dominant difference of the pion valence quark distributions at scale $\mu^2 =$ 27 GeV$^2$ is given by G19-Set4$-$G10-Set1. 
These are followed by the PDF difference results for the pion valence quark distribution at scale $\mu^2 =$ 4 GeV$^2$ for the corresponding sets. We also found that the flavor-mixing effects, which are proportional to the quark effective masses and their differences, on the PDFs are the same for both scales $\mu^2 =$ 4 and 27 GeV$^2$.

PDFs are relatively insensitive to small variations in the effective quark masses. Consequently, the shifts induced by flavor-mixing are expected to be subtle, although they may become identifiable with sufficiently precise experimental data.
However, it certainly improves the pion PDF.

For the kaon valence up-quark distributions at scale $\mu^2 =$ 27 GeV$^2$, we found that G19-Set4$-$G10-Set1 and G19-Set1$-$G10-Set4 give significant residual results. These residual values are followed by the kaon valence antistrange-quark distributions at scale $\mu^2 =$ 4 GeV$^2$ with higher magnitude.

At scale $\mu^2 =$ 27 GeV$^2$, we found that G19-Set1$-$G10-Set4 gives the dominant residual results for the kaon valence antistrange-quark distributions. However, the larger residuals are found at $x \simeq 0.4$. Similarly, the residual results for the kaon valence antistrange-quark distributions at scale $\mu^2 =$ 4 GeV$^2$ produced the same shapes and trends but different in magnitude.

The results of the present work provide new information about the effects of the implicit flavor-mixing, as defined in Eqs.~\eqref{G-K},~\eqref{G-K2}, and ~\eqref{G-K3}, induced by vacuum polarization or gluon exchange. These implicit mixings produce subtle effects on pion and kaon PDFs that can be expected to be assessed further in light of forthcoming experimental data with higher precision and lattice QCD simulations. It is worth noting that the amplitude of the implicit mixing is proportional to the quark effective masses and their differences. Further study using a more sophisticated approach is required to investigate all mixings at once. In addition, this study provides important guidance for the lattice QCD simulation, as has been done in Ref.~\cite{Horsley:2010th}, to compute the flavor-mixing contribution (quark mass differences) in PDFs or other observables~\cite{Hutauruk:2018zfk} that might be confirmed by the EIC~\cite{Arrington:2021biu}, the EicC~\cite{Anderle:2021wcy}, the JPARC~\cite{Sawada:2016mao}, and the AMBER/COMPASS++ at CERN~\cite{Adams:2018pwt}.

Experimentally, if future data achieve a precision of approximately 1\% to 2\%, it may become possible to test whether these flavor-mixing effects are indeed present for the adopted values of the effective quark masses and the NJL-model coupling constant. On the other hand, if such mixing effects are found to be necessary for reproducing the experimental data, but with slightly different magnitudes, this could provide useful constraints on the constituent-quark effective masses and the effective quark-antiquark interaction, as characterized by the NJL-model coupling constant.

\section{Data Availability}
There are no publicly available research data or software supporting this manuscript. Requests for further information or data should be sent to the authors.

\section*{ACKNOWLEDGMENTS}
F.L.B. is a member of the INCT-FNA Proc. 408419/2024-5, and was partially supported by Grants No. CNPq-307792/2025-0 and No. CNPq-407162/2023-2. P.T.P.H. thanks the Asia Pacific Center for Theoretical Physics (APCTP), where part of this work was done during my visit. This work was supported by the World Premier International Research Center Initiative (WPI) of Hiroshima University, MEXT, Japan. P.T.P.H. was partially supported by the RCNP Collaboration Research Network Program under Project No.COREnet 057.

\bibliographystyle{elsarticle-num}
\bibliography{paperfinal}
\end{document}